\documentstyle[12pt,amssymbols]{article}
\newcommand{\bq}{\begin{equation}}
\newcommand{\eq}{\end{equation}}
\newcommand{\bqa}{\begin{eqnarray}}
\newcommand{\eqa}{\end{eqnarray}}
\newcommand{\ra}{\rightarrow}

\def\half{{1 \over 2}}

\def\D{\Delta}
\def\a{\alpha}
\def\b{\beta}
\def\g{\gamma}
\def\d{\delta}

\def\l{\lambda}

\def\ov{\over}

\def\ed{\end{document}}
\def\ws{\;\;}
\def\ra{\rightarrow}

\def\2pi{1\over 2\pi i}

\def\newline{\hfil\break}

\def\ra{\rightarrow}

\def\sq2{\sqrt{2}}
\def\sqk2{\sqrt{2(k+2}}
\def\sqk{\sqrt{k}}

\def\be{\begin{equation}}
\def\ee{\end{equation}}
\def\br{\begin{array}}
\def\er{\end{array}}
\def\bea{\begin{eqnarray}}
\def\eea{\end{eqnarray}}
\def\ba{\begin{equation}\begin{array}}
\def\ea{\end{array}\end{equation}}
\def\bac{\begin{equation}\begin{array}{rll}}

\def\qbinom#1#2{{#1}\atopwithdelims[]{#2}}
\newcommand{\uq}{U_q (\widehat{sl(2)})}
\newcommand{\ug}{U_q (\widehat{g})}

\newcommand{\un}{U_q (\widehat{sl(n)})}
\newcommand{\ut}{U_q (\widehat{sl(3)})}

\def\Z{{\Bbb Z}}

\def\qd{ {_1{{\cal{D}}}}_z }
\def\qdw{ {_1{{\cal{D}}}}_w }

\def\vt{\tilde{\Phi}}

\def\sl{\sum\limits}

\begin{document}
%\bibliographystyle{unsrt}
%%%%%%%%%%%%%%%%%%%%%%%%%%%%%%%%%%%%%%%%%%%%%%%%%%%%%%%%%
%% TITLE PAGE
\begin{titlepage}
\rightline{CRM-2166}
\rightline{hep-th/9404021}
\rightline{Mar 31, 1994}
\vbox{\vspace{15mm}}
\vspace{1.0truecm}
\begin{center}
{\LARGE \bf A Free-field Representation of the Screening
Currents of $U_q(\widehat{sl(3)})$
}\\[8mm]
{\large A.H. BOUGOURZI$^{1}$  and  ROBERT A. WESTON$^{2}$}\\
[3mm]{\it Centre de Recherche Math\'ematiques,
Universit\'e de Montr\'eal\\
C.P. 6128-A, Montr\'eal (Qu\'ebec) H3C 3J7, Canada.}\\[15mm]
\end{center}
\begin{abstract}
We construct  five independent screening currents
associated with the $U_q(\widehat{sl(3)})$ quantum current algebra.
The
 screening currents
are expressed as exponentials
of the eight basic deformed bosonic fields that are
required in the quantum analogue of the
Wakimoto realization of the current algebra.
Four of the screening currents are `simple', in that each one
is given as a single exponential field.
 The fifth is
expressed as an
infinite sum of exponential fields.
For reasons we discuss, we expect that the structure
of the screening currents
for a general quantum affine algebra will be similar to the $\ut$
case.
\end{abstract}
\footnotetext[1]{
Email: {\tt bougourz@ere.umontreal.ca}
}
\footnotetext[2]{
Email: {\tt westonr@ere.umontreal.ca}
}
\end{titlepage}
\section{Introduction}
It has recently been realized that
quantum affine algebras play the role of dynamical symmetries
in many two-dimensional quantum integrable systems.
This is the case in both continuum theories such as the
Sine-Gordan model \cite{BeLe92,BeLe93}, and in lattice models such as the
$XXZ$ quantum spin chain \cite{Daval92,collin}.
For the spin-$\half$ XXZ chain (the anisotropic
quantum Heisenberg model), the dynamical symmetry is that
of $\uq$ at level 1.
The calculation of equal-time correlation functions of local
operators in this model  turns into the problem of
calculating traces of the vertex operator interwiners
defined in \cite{FrRe92,Daval92} as
\be \vt_{\lambda}^{\mu V}(z):V(\lambda)\ra V(\mu)\otimes V_z .\ee
Here $V(\lambda)$ and $V(\mu)$ are the irreducible highest weight
representations of $\uq$ associated with the dominant integral
weights $\l$ and $\mu$, and $V_z \simeq V\otimes {\Bbb C}[z,z^{-1}]$ is
the evaluation representation.
This analysis of the $XXZ$ model has been generalized, both to higher spin $s$
\cite{BoWe93b,Idz93}, in which case the dynamical symmetry is $\uq$ at
level $k=2s$, and to $sl(n)$ spin chains \cite{Koy93}, where the symmetry is
$\un$.

The desired traces may be obtained
via two different approaches. Firstly, it is possible
to derive quantum difference equation obeyed by these
objects \cite{Jimal93b,Fodal93}.
 However, generically, these equations will
have an infinite number of solutions, and some additional
information about the expected analytic structure of
the `true' solution is required in order to isolate it.
The second approach is to use the free field representation
of the algebra and vertex operators -  where it exists.
This approach yields an integral expression
\cite{collin,Idz93,BoWe93b,Koy93},
which in some simple cases can be integrated explicitly,
and in other cases yield enough analytic information in
order to select
the correct solution of the difference equation for the
same quantity \cite{Koy93}.

The main problem associated with the free field
realization is that in general the Fock space constructed
from the free fields is
completely reducible, i.e., it contains infinitely many
null states.  The unique irreducible subspace, which is embedded in this large
Fock space, and  over which one
should calculate the traces,
can be obtained by modding out
the null states and their descendants \cite{FeFr90,Malal86}.
This problem has been resolved
in the context of conformal field theory through the
use of screening current operators
\cite{DoFa84,Dot91,BeFe90}.
The main property of a screening current is that
its operator product expansions (OPEs) with the
current algebra under consideration are
either regular or total derivatives.

In the context of conformal field theories
with Kac-Moody symmetry,  it is known that the
screening currents split into two different sets \cite{Boual91}.
The screening currents  in the first
set are each expressed as
a finite sum of products of two terms. The first term consists of a first
derivative
of the basic free bosonic fields which arise
in the Wakimoto realization of the current algebra
\cite{Wak86}.
The second term  is expressed as an
exponential of these same basic fields.
The screening currents in this first set may be constructed in
an algorithmic way from the bosonized expressions for
the raising step currents \cite{Boual90}.
The screening currents in the second set are
expressed  only  as exponentials
of the basic fields;  they do not depend on their derivatives
\cite{Boual91}.
These screening currents are now related to both
raising and lowering step currents,
and there is the additional complication that some
of them involve
infinite sums of such exponentials.

In this paper we are interested in the case of quantum
affine algebras.
It has recently been shown in Ref.
\cite{Awaal93} that
for  $\un$ there exists
 an analogue of the  first
set of screening currents described above.
The screening currents now depend on both free deformed bosonic
fields and their  first quantum derivatives,
 and again can be read off from the bosonization
of the raising step currents of the quantum current
algebra. $\un$ is presently the only algebra
for which a Wakimoto
realization in terms of free deformed bosons is
available \cite{Awaal93}.
The purpose of this paper is to show that the second set
of screening currents
also has a quantum analogue. We choose
the special case of $\ut$
because in the classical case $\widehat{sl(3)}$ is
the simplest example which embodies all
the features of $\hat{g}$:
the screening currents which
involve infinite sums  appear here but not
in $\widehat{sl(2)}$ (for  $\widehat{sl(2)}$ only the
screening currents which involve single
exponentials appear).
For the `classical case' of $\widehat{sl(n)}$,
no more new features, other than computational complexity,
arise in explicit computation of the screening currents than are present for
$\widehat{sl(3)}$  \cite{Boual91}.
It is legitimate to expect that the
$\ut$  current algebra will also maintain its
special  status in the quantum case.
 Nevertheless, we will
briefly discuss  the screening currents in the general case
of $\un$.

The paper is organized as follows. In Section 2, we
review the bosonization of the $\ut$ quantum
current algebra at general level $k$.
In section 3,
we present in four successive steps
our method for constructing the second set
of  screening currents associated with this algebra.
In section 4, by way of comparison, we discuss
the simpler case of $\uq$. In addition
we comment on the extension of our method to
$\un$ and  make a conjecture concerning the single exponential
screening currents. Section 5 is devoted to our conclusions.
In appendix A, we give the definition of a general
quantum current algebra $\ug$. In appendix
B, we present four tables of OPEs corresponding to the
four steps of section 3.
%%%%%%%%%%%%%%%%%%%%%%%%%%%%%
%%%%%%%%%%%%%%%%%%%%%%%%%%%%%%
\section{Bosonization of the $\ut$ Current Algebra}
The definition of a quantum affine algebra was given
in references \cite{Dri85,Dri86,Jim85}.
For a brief summary of this definition, and that
of the associated quantum current algebra, see Appendix
\ref{app1} (more details are given in reference
\cite{Awaal93}).
 Here we review the bosonization (the
Wakimoto realization) of
the quantum current algebra of $\ut$.

The number of deformed bosonic fields required for the
free-field representation of the $\un$ current algebra is
equal to the dimension of $sl(n)$, i.e., $n^2-1$. These fields
may all be expressed in terms of the following fundamental
$d^i$ fields
$(i=1,\cdots,n^2-1)$ \cite{Awaal93}:
\bac
d^i(M,N|z,\a)&\equiv &{M\ov N}d^i+{M\ov N}d^i_o\ln(z)-\sl_{n\ne 0}
{ q^{-\a |n|}[Mn] \ov [Nn][n]} d^i_nz^{-n}, \\
d^i(M|z,\a)&\equiv &d^i(M,1|z,\a)=Md^i+Md^i_0\ln(z)-\sl_{n\ne 0}
{ q^{-\a |n|}[Mn] \ov [n]^2} d^i_nz^{-n}, \\
d^i(z,\a)&\equiv &d^i(1,1|z,\a)=d^i+d^i_0\ln(z)-\sl_{n\ne 0}
{ q^{-\a |n|}\ov [n]} d^i_nz^{-n},\\
d^i_\pm(z)&\equiv&d^i(zq^{\mp(\alpha-1)},\alpha)-
d^i(zq^{\mp\alpha},\alpha-1)
=  \pm d^i_0\ln q \pm (q-q^{-1})\sum_{\pm n>0}
d^i_nz^{-n}.\ea
These fields satisfy the following useful identities:
\bac
d^i(M,N|zq^\beta,\alpha)+d^i(M^\prime,N|zq^\gamma,
\alpha)&=&d^i(M+M^\prime,N|zq^\delta,\alpha),\\
{\rm with}\quad
\delta&=&
\left\{ \begin{array}{ll}
\gamma+M, & {\rm if}\ws M+M^\prime=\beta-\gamma,\\
\gamma-M, & {\rm if}\ws M+M^\prime=\gamma-\beta, \end{array}
 \right.\\
d^i(wq^\beta,\alpha)+d^i(\beta|wq^{-1},\alpha)
&=&d^i(wq^{-\beta},\alpha)+d^i(\beta|wq,\alpha)
\label{ids}. \ea
Let us recall that the OPE of the exponentials
of two fields $d^i(M|z,\a)$ and $d^j(M^{\prime}|w,\b)$
 is given by
\be e^{d^i(M|z,\a)} e^{d^j(M^{\prime}|w,\b)}=
e^{<d^i(M|z,\a) d^j(M^{\prime}|w,\b)>}
:e^{d^i(M|z,\a)+d^j(M^{\prime}|w,\b)}:,\ee
where as usual the normal ordering symbol :\dots : means that
the annihilation modes $\{d^i_n, n\geq 0\}$ are moved
to the right of the creation modes $\{d^j_{n}, n<0 \}$ and
of the
shift mode $d^j$.
For later purposes, we list the three generic vacuum expectation
values that will be required:
\bac
<d^i(M|z,\alpha)d^j(M^\prime|w,\beta)>&=&
 M M^{\prime}[d_0^i,d^j] -\sl_{n>0} q^{-(\alpha+\b) n}
{[d^i_n,d^j_{-n}][Mn][M^{\prime}n] \over
[n]^4}z^{-n}w^n,\\
<d^i_+(z)d^j(M|w,\alpha)>&=&
 M[d_0^i,d^j]\ln q+(q-q^{-1})\sl_{n>0}q^{-\alpha n}{[d_n^i,
d_{-n}^j][Mn]\over
[n]^2}z^{-n}w^n,\\
<d_-^i(z)d^j(M|w,\alpha)>&=&-M[d_0^i,d^j]\ln q.
\ea
In what follows, all simple operators or products
of operators defined at the  points $zq^n$ with the same
point $z$, for some integers
$n$,  are understood to be normal ordered. The q-integer
$[n]$ is as usual defined by
\be
[n]={q^n-q^{-n}\over q-q^{-1}}.
\ee

For $\ut$, there are thus 8 bosonic fields labeled as
 $a^i\equiv d^i,i\equiv 1,2;\ws b^i\equiv d^{i+2},i=1,2,3;
\ws
c^i\equiv d^{i+5},i=1,2,3.$
The corresponding deformed Heisenberg algebras are
\cite{Awaal93}
\ba{lllll}
[a_n^i,a_m^j]&={1\ov n} [n(k+3)][a^{ij}n]\d^{n+m,0},\quad &[a_0^i,a^j]&
=(k+3)a^{ij},&\quad i,j=1,2;\\
{[}b_n^i,b_m^j]&=-{1\ov n}[n]^2\d^{n+m,0}\d^{ij},\quad &[b_0^i,b^j]&
=-\d^{ij},&\quad i,j=1,2,3;\\
{[}c_n^i,c_m^j]&={1\ov n} [n]^2 \d^{n+m,0}\d^{ij},\quad &[c_0^i,c^j]&=\d^{ij},
&\quad i,j=1,2,3;\label{heis}\ea
where $a^{ij}$ is the Cartan matrix of $sl(3)$, i.e.,
\be (a^{ij})=\pmatrix{ 2& -1\cr -1&2 }.
\label{fields}
\ee
All commutators of different letters vanish, and
$k\in \Z$ is called
the level of the algebra.
We use the identities \ref{ids} to re-express
the bosonization of the  current generators of $\ut$ given in ref.
\cite{Awaal93}
in a suitable
form for our purposes:
\bac
E^{-,1}(z)&=&J^1(z)+J^2(z),\\
E^{-,2}(z)&=&J^3(z)-J^4(z),\\
E^{+,1}(z)&=& -J^5(z), \\
E^{+,2}(z)&=&-J^6(z)-J^7(z),\\
\psi^i_\pm(zq^{\pm k/2})&=&e^{\varphi^i_\pm(z)},
\label{currents}\ea
where,
\be
J^i(z)=e^{X^i(z)} \qd e^{Y^i(z)}, \ws i=1,\cdots,7;
\ee
and
\bac
X^1(z)&=&-a^1(z,(k+1)/2)-b^1(k+1|z,1)+b^2(z,k+1)-b^3(z,k+2)-c^1(k|z,0),\\
X^2(z)&=&a^1(zq^{(k+3)/2},1)-a^1(zq^{(k+1)/2},0)
-b^2(zq^{k+3},0)+b^3(zq^{k+3},1)+c^3(zq^{k+2},0),\\
X^3(z)&=&-a^2(z,(k+1)/2)-b^2(k+2|z,1)-c^2(k+1|z,0),\\
X^4(z)&=&a^2(zq^{-(k+3)/2},1)-a^2(zq^{-(k+1)/2},0)
-b^1(zq^{-(k+1)},1)+b^2(z^{-(k+3)},1)\\
&&-b^2(zq^{-(k+1)},-1)+
b^3(zq^{-(k+1)},0)+c^3(zq^{-(k+1)},0),\\
X^{5}(z)&=&-b^1(z,-1),\\
X^{6}(z)&=&-b^2(qz,-1)+b^3(qz,0)-b^3(z,-1)-b^1(q^2z,0)+b^1(qz,-1),\\
X^7(z)&=&b^1(z,0)+c^1(z,0)-b^3(z,-1),\\
Y^1(z)&=&a^1(z,(k+3)/2,0)+b^1(k+2|z,1)-b^2(z,k+2)+b^3(z,k+3)+c^1(k+1|z,0),\\
Y^2(z)&=& -c^2(zq^{k+2},0) ,\\
Y^3(z)&=&a^2(z,(k+3)/2)+b^2(k+3|z,1)+c^2(k+2|z,0),\\
Y^4(z)&=&-c^1(z^{-(k+1)},0),\\
Y^5(z)&=& -c^1(z,0),\\
Y^{6}(z)&=&-c^2(qz,0),\\
Y^7(z)&=&-c^3(z,0),\\
\varphi^1_\pm(z)&=&a^1_\pm(zq^{\pm(k+3)/2})+b^1_\pm(zq^{\pm k})+b^1_\pm
(zq^{\pm(k+2)})-b^2_\pm(zq^{\pm(k+2)})+b^3_\pm(zq^{\pm(k+3)}),\\
\varphi^2_\pm(z)&=&a^2_\pm(zq^{\pm(k+3)/2})-b^1_\pm(zq^{\pm (k+1)})+b^2_\pm
(zq^{\pm(k+1)})+b^2_\pm(zq^{\pm(k+3)})+b^3_\pm(zq^{\pm k}).
\ea
Here the quantum derivative $\qdw$ appears. It is
defined more generally  by
\be
{ {_p{{\cal{D}}}}_w }f(z)={ {f(zq^p)-f(zq^{-p}) }\over
 {z(q-q^{-1})}},\quad p\in {\Bbb Z}/\{ 0\}.
\ee
For latter use, we also introduce the notation,
\be
J^i(z)={ J_+^i(z)-J_-^i(z) \ov {z(q-q^{-1})}},\ws\ws
 {\rm with} \ws
J_{\pm}^i(z)=e^{X_i(z)} e^{Y_i(zq^{\pm 1})}.\ee
%%%%%%%%
\newpage
\section{The Screening Currents of $\ut$\label{sl3}}
The screening currents are constructed by carrying out the
following sequence of steps.
\subsubsection*{Step 1}
Let $J(z)=e^{X(z)}\qd e^{Y(z)}$ be any of the seven basic currents
defining the step currents $E^{\pm ,i}(z), i=1,2$
of equation \ref{currents}, and let $G(w)=e^{g(w)}$ be a
candidate screening current constructed  such
that its OPE with $J(z)$ is the following total quantum derivative:
\be
J(z)G(w)=
{q^\a \ov  {z(q-q^{-1}) } }
\left(
{{q^\b J_+(z)G(w)}\ov {z-wq^\beta}}
- { {q^{-\beta} J_-(z)G(w)}\ov {z-wq^{-\beta}}}
\right)
\sim q^\alpha\;  {_\b{\cal D}}_w\left(
{f(w)\over z-w}\right),
\label{cond2}\ee
where $\sim$ means `equal up to
regular terms', $\a$ and $\b$ are integers to be determined, and   $f(w)$
satisfies the condition
\be
f(w)=:J_+(w)G(wq^{-\beta}):=:J_-(w)G(wq^{\beta}):.\ee
This condition translates into the following relation:
\be
Y(wq)+g(wq^{-\beta})=Y(wq^{-1})+g(wq^\beta).\ee
There are two obvious solutions to this  latter equation:
\ba{llllll}
\beta &=&1 \ws &{\rm and}\ws g(w)&=&Y(w),\\
\beta &=-&1 \ws &{\rm and}\ws g(w)&=-&Y(w).\label{bcond}\ea
However, for the OPE of $G(w)=e^{\beta Y(w)}$ with $J(z)$
to then be a total derivative,
we also require the following:
\be
J_{\pm}(z)e^{ Y(w)}=\left\{\begin{array}{lll}
{q^{\alpha\pm 1}\over
 z-wq^{\pm 1}}:J_\pm(z)e^{ Y(w)}:,\quad &{\rm for}\ws &
\beta=1,\\
q^{-\alpha\pm 1}( z-wq^{\mp 1}):J_\pm(z)e^{Y(w)}:,\quad &{\rm for}\ws &\b=-1.
\end{array}\right.
\label{cond3}\ee
In Table 1 we compute all the OPEs of the form $J^i(z) e^{Y^i(w)},\> i=1,
\dots , 7$.  Examining this table and relation \ref{cond3}, one can
see that it is possible to satisfy
condition \ref{cond2} if we choose $G^i(z)=
e^{\b^i Y^i(z)} $ for all $i=1,\cdots,7$, with the
$\b^i$ given by
\be \beta^i = \left\{ \begin{array}{ll}
-1, & i=2,4,5,6,7\\
+1, & i=1,3 \end{array} \right.. \ee
Note that one can also read off the value $\alpha^i$ appearing in each OPE
 $J^i(z) G^i(w)$ from Table 1. We find
\be
\alpha^i=
\left\{ \begin{array}{ll}
0, & i=1,3\\
-k-2, & i=2\\
k+1, &i=4\\
0,&i=5,7\\
-1&i=6 \end{array} \right.. \ee

\subsubsection*{Step 2}
For any of the $G^j(w),\> j=1,\dots ,7$, to be  a genuine screening current,
its
OPEs  with all the remaining
currents, i.e.,  $J^i(z)G^j(w)$, $i\neq j$ and $\psi^i_\pm(zq^{\pm
k/2})G^j(w)$,
must be either  regular or total quantum derivatives.
We give these OPEs in Table 2.
They are indeed all regular
except for the OPE $J^2(z)G^1(w)$, which  is neither regular
nor a total derivative. Therefore $G^1(w)=e^{Y^1(w)}$ fails to be
a genuine screening current.
It might at this stage appear that we have already
constructed six screening currents. However, as
$G^2(z)=G^6(zq^{k+1})$ and $G^4(z)=G^5(zq^{-k-1})$, there are
only four independent screening currents (it is trivial to see that
if $G(w)$ is a screening current, then so
is $G(wq^n), n\in {\Bbb Z}$). We denote these currents by
\bac S^2(w)&=&G^2(w),\\
S^3(w)&=&G^3(w),\\
S^4(w)&=&G^4(w),\\
 S^7(w)&=&G^7(w).\label{4sc}\ea

We have shown that from each basic current $J^i(z), i=2,\dots, 7,$
one can
construct a screening current $S^i(w)$. Putting it another
way - the number of screening
currents one can construct from each current $E^{\pm,i}(z)
,\> i\neq 1$, is equal to the number
of $J^i(z)$ it contains. This rule fails for $E^{-,1}(z)=J^1(z)+J^2(z)$:
it has so far
led to only one screening current $S^2(w)$ (the one associated with $J^2(z)$)
instead of two.
 This is because  $J^1(z)$
fails to yield a screening current on its own.
 However,  we show in the next step that all is not lost;
this failure can still be corrected in a nontrivial way with the
help of $J^2(z)$.
%%%%%%%%%%%%%%%%%%%%%%%%%%%%%%%%%%%%%%%%%%%%%%%%%%%%
 \subsubsection*{Step 3}
%%%%%%%%%%%%%%%%%%%%%%%%%%%%%%%%%%%%%%%%%%%%%%%%%%%%
Our aim in this section is to extract a second candidate screening
current $G^{(1,2)}(w)$
from
$E^{-,1}(z)$ by using both $J^1(z)$ and $J^2(z)$ simultaneously.
To this end, we compute the OPEs involving $J^i_\pm(z)e^{Y^j(w)}; i,j=1,2$
and $J^i_\pm(z)e^{X^1(w)-X^2(w)}; i=1,2$. These OPEs are presented in
Table 3.
As discussed above, the OPE $J^1_\pm(z)e^{Y^1(w)}$ is a
total quantum derivative, i.e., $e^{Y^1(w)}$ is a screening current with
respect to $J^1(z)$; the problem is that the
OPE $J^2_\pm(z)e^{Y^1(w)}$ is singular with poles
at $z=wq^{\pm 1}$. The idea is to correct this
problem using a recursive technique.
This is achieved as follows. First note
 that according to Table 3 the OPEs of the operator
\be
G^{(1,2)}_0(w)\equiv e^{Y^1(w)+Y^2(w)}
\ee
with $J^i(z)$, $i=1,2$, though neither regular nor
total quantum derivative, have the following `nice' forms:
\bac
J^1(z)G^{(1,2)}_0(w)&=&e^{Y^2(w)}\qdw \left(
{e^{X^1(w)+Y^1(wq)+Y^1(wq^{-1})}\over
z-w}\right),\\
J^2(z)G^{(1,2)}_0(w)&=&q^{k+2}e^{Y^1(w)}\qdw
\left({e^{X^2(w)+Y^2(wq)+Y^2(wq^{-1})}\over
z-w}\right)
.\label{OPE3}\ea
The relevance of these nice forms stems from the
{\it two} possible quantum analogues of the chain rule:
\bea
\qdw \left(f(w)h(w)\right)&=&h(wq)\qdw f(w)+f(wq^{-1})\qdw
h(w),\label{crule1}\\
\qdw \left(f(w)h(w)\right)&=&h(wq^{-1})\qdw f(w)+f(wq)\qdw h(w).
\label{crule2}\eea
To appreciate the r\^ole of these two different rules, let us consider the
OPE $J^1(z)G^{(1,2)}_0(w)$ of equation \ref{OPE3}.
If we construct two operators,
which we denote by
$q^{\alpha_{-1}^{m_{-1}}}G^{(1,2),m_{-1}}_{-1}(w)$, where
$m_{-1}=0,1$ and
$\alpha_{-1}^{m_{-1}} \in {\Bbb Z}$, such that
\be
J^2(z)q^{\alpha_{-1}^{m_{-1}}}
G^{(1,2),m_{-1}}_{-1}(w)
=
{
{ e^{X^1(wq^{2m_{-1}-1})+
Y^1(wq^{2m_{-1}})
+Y^1(wq^{2m_{-1}-2})}}
\over
z-wq^{2m_{-1}-1}
}
\;\qdw e^{Y^2(wq^{2m_{-1}-1})}
\label{req1},\ee
then the following OPEs add up to a total
quantum derivative:
\be J^1(z)G^{(1,2)}_{0}(w)
+J^2(z)q^{\alpha_{-1}^{m_{-1}}}
G^{(1,2),m_{-1}}_{-1}(w)=
 \qdw\left(
{e^{X^1(w)+Y^1(wq)
+Y^1(wq^{-1})+Y^2(wq^{2m_{-1}-1})}\over
z-w}\right).
\label{J1J2}\ee
The two values, $m_{-1}=0,1$, correspond  to using \ref{crule1} and
\ref{crule2} respectively for the rhs.
Here
we use the convention that
$G^{(1,2),m_{n}}_{n}(w)$  (that will be introduced
shortly for general $n\in \Z$)
are pure exponential operators with no overall multiplicative factors.
In this notation,
$G^{(1,2)}_{0}(w)$ is identified with
$G^{(1,2),m_{0}=0}_{0}(w)$, with $\alpha_0^{m_0=0}
=0$.
{}From the requirement \ref{req1} and Table 3 we find that
the operator $q^{\alpha_{-1}^{m_{-1}}}G^{(1,2),m_{-1}}_{-1}(w)$  must be given
by
\bac
G^{(1,2),m_{-1}}_{-1}(w)&=&
e^{X^1(wq^{2m_{-1}-1})-
X^2(wq^{m_{-1}-1})+Y^1(wq^{2m_{-1}})+
Y^1(wq^{2m_{-1}-2}},\ws m_{-1}=0,1; \ws {\rm and},\\
\alpha_{-1}^{m_{-1}}&=&2m_{-1}-k-5.\ea
{}From the preceeding discussion,
a natural candidate for our screening  is therefore
$G^{(1,2)}(w)=q^{\alpha_{0}^{m_{0}=0}}
G^{(1,2),m_{0}}_{0}(w)+
q^{\alpha_{-1}^{m_{-1}}}
G^{(1,2),m_{-1}}_{-1}(w)$.
However, using Table 3 one finds that the crossing terms
$J^2(z)q^{\alpha_{0}^{m_{0}=0}}G^{(1,2),m_{0}}_{0}(w)$
and
$J^1(z)q^{\alpha_{-1}^{m_{-1}}}G^{(1,2),m_{-1}}_{-1}(w)$
in the OPE $E^{1,-}(z)G^{(1,2)}(w)$ do not add up to be
either
regular or a total quantum derivative. They are
given respectively by \ref{OPE3} and
\bac
J^1(z)q^{\alpha_{-1}^{m_{-1}}}
G^{(1,2),m_{-1}}_{-1}(w)&=&
q^{\alpha_{-1}^{m_{-1}}-2m_{-1}+1}e^{-X^2(q^{2m_{-1}-1})}
\qdw\left({e^{h(w)}\over z-wq^{2m_{-1}}}\right)\\
&& +q^{\alpha_{-1}^{m_{-1}}-1}
{
e^{
h(wq^{-1})-X^2(wq^{2m_{-1}-1})}
\over (z-wq^{2m_{-1}
-1})(z-wq^{2m_{-1}-3}) },
\ea
with
\be
h(w)=X^1(wq^{2m_{-1}})+X^1(wq^{2m_{-1}-2})+Y^1(wq^{2m_{-1}+1})+
Y^1(wq^{2m_{-1}-1})+Y^1(wq^{2m_{-1}-3}).
\ee
This means that one has to  correct the above operator
$G^{(1,2)}(w)$ by adding in two more terms, denoted by
$q^{\alpha_{1}^{m_{1}}}G^{(1,2),m_{1}}_{1}(w)$ and
$q^{\alpha_{-2}^{m_{-2}}}G^{(1,2),m_{-2}}_{-2}(w)$, \thinspace
 such that
the OPEs $J^1(z)q^{\alpha_{1}^{m_{1}}}G^{(1,2),m_{1}}_{1}(w)$
$+J^2(z)q^{\alpha_{0}^{m_{0}}}G^{(1,2),m_{0}}_{0}(w)$ and
$J^1(z)q^{\alpha_{-1}^{m_{-1}}}G^{(1,2),m_{-1}}_{-1}(w)
+J^2(z)q^{\alpha_{-2}^{m_{-2}}}G^{(1,2),m_{-2}}_{-2}(w)$ are
total quantum derivatives.

This process continues infinitely many
times. Table 3 allows us to perform the
appropriate corrections recursively to end up with the following candidate
screening current, which is expressed as
a series,
\be
G^{(1,2)}(w)=\sl_{n\in Z} q^{\alpha_{n}^{m_{n}}}G^{(1,2),m_{n}}_{n}(w),
\label{g12}\ee
where
\bac
G^{(1,2),m_{n}}_{n}(w)
&=&\exp\left(
-\sl_{i=1}^{n}
\left(X^1(wq^{2m_n-2i+1})
-X^2(wq^{ 2m_n-2i+1})\right)\right.\\
&&-\left.\sl_{i=1}^{n-1}Y^1(wq^{2m_n-2i})
+\sl_{i=0}^{n}Y^2(wq^{2m_n-2i})\right),\quad n>0,\\
G^{(1,2),m_{-n}}_{-n}(w)
&=&\exp\left(\sl_{i=1}^{n}\left(
X^1(wq^{ 2m_{-n}-2i+1})
-X^1(wq^{ 2m_{-n}-2i+1})\right)\right.\\
&&+\left.\sl_{i=0}^{n}Y^1(wq^{2m_{-n}-2i})
-\sl_{i=1}^{n-1}Y^2(wq^{2m_{-n}-2i})\right),\quad n>0,\\
G^{(1,2),m_{0}}_{0}(w)
&=&\exp\left(({Y^1(w)+Y^2(w)}\right),
\ea
and
\ba{llll}
\alpha_{n}^{m_{n}}&=&-n^2+n(k+2)+2m_n,\quad &n>0,\\
\alpha_{-n}^{m_{-n}}&=&-n^2-n(k+4)+2m_{-n},\quad &n>0,\\
\alpha_{0}^{m_{0}}&=&0.
\ea
For $n>0$,\,
$m_{\pm n}$ is such that $0\leq m_{\pm n}\leq n$,
 and is defined recursively from
$m_0=0$ by  either $m_{\pm n}=m_{\pm (n-1)}$ or
$m_{\pm n}=m_{\pm (n-1)}+1$,
depending on whether we use \ref{crule1} or \ref{crule2}
 respectively to construct $G^{(1,2),m_{\pm n}}_{\pm n}(w)$ from
$G^{(1,2),m_{\pm (n-1)}}_{\pm(n-1)}(w)$. In the former case, i.e.,
 $m_{\pm n}=m_{\pm (n-1)}, \; n>0$,\  one can easily check that
\bac
&&J^1(z)q^{\alpha_{n}^{m_{n}}}G^{(1,2),m_{n}}_{n}(w)
+J^2(z)q^{\alpha_{n-1}^{m_{n-1}}}G^{(1,2),m_{n-1}}_{n-1}(w)\\
&&=
q^{\alpha_{n}^{m_n}+n-2m_n}\qdw \left({F(w)\over
z-wq^{2m_n}}\right),\quad n\geq 1,\\
&&J^1(z)q^{\alpha_{n}^{m_{n}}}G^{(1,2),m_{n}}_{n}(w)
+J^2(z)q^{\alpha_{n-1}^{m_{n-1}}}G^{(1,2),m_{n-1}}_{n-1}(w)\\
&&=
q^{\alpha_{n}^{m_n}-n-2m_n}\qdw \left({H(w)\over
z-wq^{2m_n}}\right),\quad n\leq 0,
\ea
where,
\bac
F(w)&=&:J^1_-(wq^{2m_n})G^{(1,2),m_n}_n(wq):=
:J^2_+(wq^{2m_n})G^{(1,2),m_{n-1}}_{n-1}(wq^{-1}):,\\
H(w)&=&:J^1_+(wq^{2m_n})G^{(1,2),m_n}_n(wq^{-1}):=
:J^2_-(wq^{2m_n})G^{(1,2),m_{n-1}}_{n-1}(wq):,\\
\alpha^{m_n}_n&=&\alpha^{m_{n-1}}_{n-1}-2n+k+3, \quad n\geq 1,\\
\alpha^{m_n}_n&=&\alpha^{m_{n-1}}_{n-1}-2n+k+5, \quad n\leq 0.
\ea
In the latter case, i.e., $m_{\pm n}=m_{\pm (n-1)}+1,\; n>0
$, \ we find
\bac
&&J^1(z)q^{\alpha_{n}^{m_{n}}}G^{(1,2),m_{n}}_{n}(w)
+J^2(z)q^{\alpha_{n-1}^{m_{n-1}}}G^{(1,2),m_{n-1}}_{n-1}(w)\\
&&=
q^{\alpha_{n}^{m_n}+n-2m_n}\qdw \left({F^\prime(w)\over
z-wq^{2m_n-2n}}\right),\quad n\geq 1,\\
&&J^1(z)q^{\alpha_{n}^{m_{n}}}G^{(1,2),m_{n}}_{n}(w)
+J^2
(z)q^{\alpha_{n-1}^{m_{n-1}}}G^{(1,2),m_{n-1}}_{n-1}(w)\\
&&=
q^{\alpha_{n}^{m_n}-n-2m_n}\qdw \left({H^\prime(w)\over
z-wq^{2m_n+2n}}\right),\quad n\leq 0,
\ea
where,
\bac
F^\prime(w)&=&:J^1_+(wq^{2m_n-2n})G^{(1,2),m_n}_n(wq^{-1}):=
:J^2_-(wq^{2m_n-2n})G^{(1,2),m_{n-1}}_{n-1}(wq):,\\
H^\prime(w)&=&:J^1_-(wq^{2m_n+2n})G^{(1,2),m_n}_n(wq):=
:J^2_+(wq^{2m_n+2n})G^{(1,2),m_{n-1}}_{n-1}(wq^{-1}):,\\
\alpha^{m_n}_n&=&\alpha^{m_{n-1}}_{n-1}-2n+k+5, \quad n\geq 1,\\
\alpha^{m_n}_n&=&\alpha^{m_{n-1}}_{n-1}-2n+k+3, \quad n\leq 0.
\ea
{}From the above relations it is therefore clear that the
OPE $E^{1,-}(z)G^{(1,2)}(w)$  (with $G^{(1,2)}(w)$
given by \ref{g12}) is a sum of an infinite number of quantum
total derivative terms. The sum is also a total quantum derivative
since the quantum derivative is linear.

There are an infinite number of ways of choosing the
$\{ m_n\}$ and hence an
infinite number of screening current candidates.
However $G^{(1,2)}(w)$ does not yet qualify as
a genuine screening current -  we must also
make sure that its OPEs with the remaining currents
$E^{2,-}(z), E^{i,+}(z), \psi^i_\pm(zq^{\pm k/2}); i=1,2$, are
 regular or quantum total derivatives.
This is the subject of the fourth and final step.
\subsection*{Step 4}
In order to check the consistency of $G^{(12)}(z)$
with the remaining currents
$E^{2,-}(z), E^{i,+}(z),$ $ \psi^i_\pm(zq^{\pm k/2});$
$ i=1,2$, we
construct Table 4. This table consists of all elementary
\linebreak
OPEs $J^i(z)e^{X^1(w)-X^2(w)}$, $J^i(z)e^{Y^1(w)}$,
 $J^i(z)e^{Y^2(w)}$; $i=3,\dots,7$, \
  $\psi^i_\pm(zq^{\pm k/2})
e^{X^1(w)-X^2(w)}$,\linebreak $\psi^i_\pm(zq^{\pm k/2})e^{Y^1(w)}$
 and
$\psi^i_\pm(zq^{\pm k/2})e^{Y^2(w)}$; $i=1,2.$
{}From this table one can easily show
that the relation
\be
J^3(z)q^{\alpha_{n+1}^{m_{n+1}}}G^{(1,2),m_{n+1}}_{n+1}(w)
-J^4(z)q^{\alpha_{n}^{m_{n}}}G^{(1,2),m_{n}}_{n}(w)\sim
{\rm regular}, \ws n\in \Z,
\ee
is true
if
the following conditions are satisfied:
\ba{llll}
&&m_n=m_{n+1},\quad &n\geq 0,\\
&&m_n=m_{n+1}+1,\quad &n<0,
\ea
Since $m_0=0$ this means that
\ba{llll}
&&m_n=0,\quad &n\geq 0,\\
&&m_n=-n,\quad &n< 0.
\ea
These two conditions uniquely select
a single candidate screening current from the infinite
set of candidates.
Now that all the parameters are fixed, all that
remains is to check that the  OPEs of this unique
screening current candidate $G^{(12)}(z)$  with the currents
$E^{1,+}(z)=-J^5(z)$, $E^{2,+}(z)=
-J^6(z)-J^7(z)$ and $\psi^i_\pm(zq^{\pm k/2})$ are regular. This is
confirmed  by the following relations:
\bac
&&\psi^i_\pm(zq^{\pm k/2})q^{\alpha_{n}^{m_{n}}}G^{(1,2),m_{n}}_{n}(w)
={\rm regular},\quad n\in \Z,\\
&&E^{1,+}(z)q^{\alpha_{n}^{m_{n}}}G^{(1,2),m_{n}}_{n}(w)
=-J^5(z)q^{\alpha_{n}^{m_{n}}}G^{(1,2),m_{n}}_{n}(w)
={\rm regular},\quad n\in \Z,\\
&&-J^6(z)q^{\alpha_{n}^{m_{n}}}G^{(1,2),m_{n}}_{n}(w)
-J^7(z)q^{\alpha_{n+1}^{m_{n+1}}}G^{(1,2),m_{n+1}}_{n+1}(w)
={\rm regular},\quad n\in \Z .
\ea
The last relation obviously means that the OPE of $G^{(12)}(z)$
 with
$E^{2,+}(z)=-J^6(z)-J^7(z)$ is also regular.
Therefore the
status of $G^{(12)}(z)$  can be
elevated to that of a  genuine screening current   which
we denote by $S^{1,2}(w)$.
%%%%%%%%%%%%%
\section{Generalization to $\un$}
%%%%%%%%%%%%
As far as the application of
our method to other quantum affine algebras
is concerned, the $\uq$
case is simplest because, unlike $\ut$, its
currents $E^\pm(z)$ and $\psi_\pm(z)$ do not involve
{\it sums} over basic fields $J(z)$.
They are given simply by
\bac
E^-(z)&=&e^{X^1(z)}\qdw e^{Y^1(w)},\\
E^+(z)&=&e^{X^2(z)}\qdw e^{Y^2(w)},\\
\psi_\pm(zq^{\pm k/2})&=&
e^{a^1_\pm
(zq^{\pm(k+2)/2})+b^1_\pm
(zq^{\pm k})+b^1_\pm(zq^{\pm(k+2)})},
\ea
where
\bac
X^1(z)&=&-a^1(z,k/2)-b^1(k+1|z,1)-c^1(k|z,0),\\
X^2(z)&=&-b^1(z,-1),\\
Y^1(z)&=&a^1(z,(k+2)/2)+b^1(k+2|z,1)+c^1(k+1|z,0),\\
Y^2(z)&=&-c^1(z,0).
\ea
Here the corresponding Heisenberg algebras are
defined by
\ba{llll}
[a_n^1,a_m^1]&={1\ov n} [n(k+2)][2n]\d^{n+m,0},\quad &[a_0^1,a^1]&
=2(k+2),\\
{[}b_n^1,b_m^1]&=-{1\ov n}[n]^2\d^{n+m,0},\quad &[b_0^1,b^1]&
=-1,\\
{[}c_n^1,c_m^1]&={1\ov n}
[n]^2 \d^{n+m,0},\quad &[c_0^1,c^1]&=1.\ea
%%%%%%%%%%%%
All the other commutation relations are trivial.
Since each of the currents $E^\pm(z)$ contains only
one term of the form $e^{X(z)}\qdw e^{Y(w)}$, it is
possible to extract the two single exponential
screening currents
$S^1(z)=e^{\beta^1Y^1(w)}$ and $S^2(z)=
e^{\beta^2Y^2(w)}$ from them.
No  screening current with an infinite number
of exponential
terms is present. Furthermore, for  $\uq$,
 the analogue
of our Table 1  leads to $\beta^{1}=1$
and $\beta^2=-1$.

%%%%%%%%%%%%%%%%
Recall that $\ut$ has the property that
  one of its
step  currents is expressed as a single term of the form
$J(z)$ defined by \ref{currents}
and the others are the sum of two such terms.
Thus our method can in
principle be extended to  other quantum affine algebras
who's
step currents are finite linear combinations
 of terms of the form $J^i(z)$. This structure of currents
is present for  $\un$ (the only quantum affine
algebra for which a Wakimoto
realization in terms of deformed bosonic fields currently
exists \cite{Awaal93}),
and very plausibly for general quantum affine algebras.
%%%%%%%%%%%%%%
Most probably, the only problem with extending our
technique will be the complexity of
 the Wakimoto realizations
of the other quantum
affine algebras.
For example the
Wakimoto realization of $\un,\;n>3$, requires
$n^2-1$ free deformed bosonic fields $a^i,i=1,\dots ,n-1$;
$b^i,i=1,\dots ,n(n-1)/2$ and $c^i,i=1,\dots ,n(n-1)/2$.
In addition
some of its step currents are sums of several basic
currents $J^i(z),i=1,\dots ,m$. This means
that corresponding
screening currents might be expressed as
$m-1$ infinite sums of exponential terms.
The maximum value of $m$ is $n-1$ and the
corresponding algebras of the above bosonic fields
are still given by \ref{fields} with 3 being replaced by $n$
(see Ref. \cite{Awaal93}
for more details).
Based on our results for
both $\uq$ and $\ut$,
we conjecture
that those screening currents of
$\un$ expressible as single exponentials
are given by
$S^i(z)=e^{c^i(z,0)},i=1,\dots ,n(n-1)/2$ and
$S(z)=e^{a^{n-1}(z, (k+n)/2)+b^{n-1}(k+n|z,1)+
c^{n-1}(k+n-1|z,0)}$. Here the
bosonic fields $a^{n-1}$, $b^{n-1}$, $c^{n-1}$ and $c^i, i=1,\dots ,n(n-1)/2$
are
identified respectively with $a^{n-1}$,
$b^{n-1,n}$, $c^{n-1,n}$ and $c^{i,j}, 1\leq i<j\leq n$
 of Ref. \cite{Awaal93}.
%%%%%%%%%%%%%%%%%%%%%%%%%%
%%%%%%%%%%%%%%%%%%%%%%%%%%
\section{Conclusions}
To summarise, we have found five independent screening
currents for $\ut$. Four of these are expressed as
single exponential operators in terms of free deformed
bosonic fields. They are given by the relations \ref{4sc}.
The fifth is more complicated and is
written as an infinite sum of
exponential operators, each of which is
given  in terms of free deformed
bosonic fields. As in the classical case, these are
all the screening currents that it is possible
to construct with
our method. It is still an open problem to show their
uniqueness.
One application of the above screening currents
in the case of both $\uq$ and $\ut$ is
to the computation of the
correlation functions of the higher spin
versions of XXZ model and their  $sl(3)$
generalisations.
\section*{Acknowledgements}
We wish to thank all members of CRM for their continuous
encouragement,
support and stimulating discussions, especially L. Vinet, Y. Saint-Aubin,
J. Letourneux, P. Winternitz and M.A. El Gradechi. The discussions we had
at the
beginning of this work with N. Hambli were very interesting.
A.H.B. thanks CRM for
providing him with funds
during the  early stages of this work,
and the CRSNG for a postdoctoral fellowship.
R.A.W thanks CRM for a postdoctoral fellowship.
\newpage
\appendix
\section{The $\ug$ Current Algebra\label{app1}}
\subsection{$\ug$}
The quantum affine algebra $\ug$ \cite{Jim85,Dri85,Dri88},
 associated with a rank $r$ Lie Algebra $g$,
 is generated by the elements,
$\{e_i^{\pm},t_i,i=0,\cdots, r\}$ such that
\bac
[t_i,t_j]&=&0,\\
t_ie_j^{\pm}t_i^{-1}&=&q^{\pm a^{ij}} e_j^{\pm},\\
{[}e_i,f_j]&=&\d_{ij}{ {t_i-t_i^{-1}} \ov {q-q^{-1}}},\\
\sl_{r=0}^{1-a^{ij}} (-1)^r {\qbinom{1-a^{ij}}{r}}
(e_i^\pm)^{1-a^{ij}} e_j^{\pm}
(e_i^\pm)^r&=&0,\ea
where $a^{ij}$ is the extended Cartan matrix of the affine algebra
$\hat{g}$, and ${\qbinom{n}{m}}
=[n]!/([m]![n-m]!)$, $[n]!=[n][n-1]\cdots [1]$
with $[n]=(q^n-q^{-n})/(q-q^{-1})$.
\subsection{The Drinfeld Realization}
Drinfeld has shown \cite{Dri86} that $\ug$ is isomorphic to
 the
  algebra
generated by $\{E_n^{\pm, i},H_{n\neq 0}^i,K_i,\gamma;i=0,
\cdots,r;n\in \Bbb{Z}\}$ \cite{Dri88}:
\bac
[K_i,H_n^j]&=&0\\
K_iE_n^{\pm,j}K_i^{-1}&=& q^{\pm a^{ij}} E_n^{\pm,j}\\
%%%%%
{[}H_n^i,H_m^j]&=&{1 \ov n} [a^{ij} n]
{ {\g^n-\g^{-n}} \ov {q-q^{-1}}}\d_{n+m,0}\\
%%%%%
{[}H_n^i,E_m^{\pm,j}]&=&\pm {1 \ov n} [a^{ij} n] \g^{\mp \half |n|}
E_{n+m}^{\pm,j}\\
%%%%%
{[}E_n^{+,i},E_m^{-,j}]&=&{\d^{ij}\ov {q-q^{-1}}}
(\g^{(n-m)/2}\psi_{+,n+m}^i-\g^{-(n-m)/2}\psi_{-,n+m}^i)\\
%%%%%
{[}E_{n+1}^{\pm,i},E_m^{\pm,j}]_{q^{\pm a^{ij}}}+
{[}E_{m+1}^{\pm,j},E_n^{\pm,i}]_{q^{\pm a^{ij}}}&=&0\\
%%%%%
{[}E_n^{\pm,i},E_n^{\pm,j}]&=&0,\ws {\rm for}\ws a^{ij}=0\\
%%%%%
{[}E_n^{\pm,i},[E_m^{\pm,i},E_\ell^{\pm,j}]_{q^{\mp 1}}]_{q^{\pm 1}}+
{[}E_m^{\pm,i},[E_n^{\pm,i},E_\ell^{\pm,j}]_{q^{\mp 1}}]_{q^{\pm 1}}
&=&0,\ws {\rm for } \ws a^{ij}=-1,\ea
where,
\bac
 \sl_{\pm n\geq 0}\psi_{\pm,n}^i z^{-n}&=&K_i^{\pm 1}
 \exp\left(\pm (q-q^{-1})\sl_{\pm n>0}
H_n^iz^{-n}\right),
%K_i&=&\exp(\half (q-q^{-1})H_0^i)\\
\label{drinf}\ea
and the q-commutator is defined by
\be
[X,Y]_{q^\alpha}\equiv XY-q^{\alpha}YX.
\ee
\subsection{The Quantum Current Algebra}
For the purpose of bosonization, it is convenient to
re-express \ref{drinf} as a current algebra \cite{Awaal93} that is generated
by the currents
\bac
%H^i(z)=\sl_{n\in \Zp}H_n^i z^{-(n+1)},\quad
E^{\pm,i}(z)&=&\sl_{n\in \Z} E_n^{\pm,i}z^{-n-1},\quad
\psi_{\pm}^i(z)= \sl_{\pm n\geq 0}\psi_{\pm,n}^i z^{-n},
\ea
such that
\bac
{[}\psi_{\pm}^i(z),\psi_{\pm}^j(w)]&=&0,\\
(z-q^{a^{ij}}\g^{-1} w)(z-q^{-a^{ij}}\g w)\psi_{+}^i(z)\psi_-^j(w)&=&
(z-q^{a^{ij}}\g w)(z-q^{-a^{ij}}\g^{-1} w)\psi_{-}^j(w)\psi_+^i(z),\\
%%%%
(z-q^{\pm a^{ij}}\g^{\mp \half} w)\psi_{+}^i(z) E^{\pm,j}(w)&=&
(q^{\pm a^{ij}} z-\g^{\mp \half} w) E^{\pm,j}(w)\psi_{+}^i(z),\\
%%%%
(z-q^{\pm a^{ij}}\g^{\mp \half} w)E^{\pm,j}(z)\psi_{-}^i(w)&=&
(q^{\pm a^{ij}} z-\g^{\mp \half} w) \psi_{-}^i(w)E^{\pm,j}(z),\\
%%%%
{[}E^{+,i}(z),E^{-,j}(w)]&=& {\d^{i,j} \ov (q-q^{-1})zw}
(\d(w\g/z)\psi_+^i(\g^{\half} w)-
\d(w/(z\g))\psi_-^i(\g^{-\half}w)),\\
%%%%
(z-q^{\pm a^{ij}} w)E^{\pm,i}(z)E^{\pm,j}(w)&=&(q^{\pm a^{ij}}z-w)
E^{\pm,j}(w)E^{\pm,i}(z),\\
%%%%
E^{\pm,i}(z)E^{\pm,j}(w)&=&E^{\pm,j}(w)E^{\pm,i}(z),\quad
a^{ij}=0,\\
%%%%
E^{\pm,i}(z_1)E^{\pm,i}(z_2)E^{\pm,j}(w)&-&
(q+q^{-1})E^{\pm,i}(z_1)E^{\pm,j}(w) E^{\pm,i}(z_2)+\\
E^{\pm,j}(w)E^{\pm,i}(z_1)E^{\pm,i}(z_2) + (z_1
\leftrightarrow  z_2)&=&0,\ws {\rm for} \ws
a^{ij}=-1,
%%%%%
\ea
where $\d(z)=\sl_{n\in \Bbb{Z}} z^n$.
For  $\ut$ there are six currents, \be
E^{+,i}(z),E^{-,i}(z),\psi^i_{\pm}(z),\quad i=1,2.\ee
The bosonization of these currents for arbitrary level and $n$
is constructed in \cite{Awaal93}.
The version given in Section 2 for $\ut$ is in a slightly different form
which is suitable for the purpose of constructing
the screening currents.
%%%%%
\newpage
\section{OPEs}
In this section, we list the OPEs relevant to the discussion
of section \ref{sl3}. The four tables refer to the OPEs
required in the four stages of the argument presented in
that section.\vspace{5mm}
%%%%%%%%%%%%%%%%%%%%%%
\\ \centerline{{\bf{Table 1}}}\\
\[\begin{array}{|lll|}\hline
J_{\pm}^i(z) e^{Y^i(w)} &=& { q^{\pm 1} \ov {z-wq^{\pm 1}} }
:J_{\pm}^i(z) e^{Y^i(w)} :,\ws i=1,3; \\
J_{\pm}^2(z) e^{Y^2(w)} &=& q^{k+2\pm 1} (z-w q^{\mp })
:J_{\pm}^2(z) e^{Y^2(w)} :,\\
J_{\pm}^4(z) e^{Y^4(w)} &=&  q^{-k-1\pm 1} (z-w q^{\mp }):
J_{\pm}^4(z) e^{Y^4(w)},:\\
J_{\pm}^i(z) e^{Y^i(w)} &=&q^{\pm 1} (z-w q^{\mp}):J_{\pm}^i(z) e^{Y^i(w)}:,
\ws i= 5,7; \\
J_{\pm}^6(z) e^{Y^6(w)} &=&q^{1 \pm 1} (z-w q^{\mp}):J_{\pm}^6(z) e^{Y^6(w)}:\\
\hline
\end{array}\]\\ \\
%%%%%%%%%%%%%%%%%%%%%
\centerline{{\bf{Table 2}}}\\
\[\begin{array}{|lll|}\hline
J_{\pm}^i(z) e^{\beta^j Y^j (w)} &=& \ws {\rm regular},\ws
{\rm where}\ws (i,j)\in \{(i,j);(i,j)\neq (2,1);i=1,\cdots,7;j=2,3,4,7\},\\
J_{\pm}^2(z) e^{Y^1 (w) } &=& {1 \ov {(z-wq)(z-wq^{-1})}} :J_{\pm}^2
e^{Y^1 (w) }:,\\
\psi^i_\pm(zq^{\pm k/2})e^{\beta^j Y^j (w)}&=&\ws {\rm regular }, \ws
i=1,2\ws j=1,\dots, 7.
\\ \hline
\end{array}\]\\ \\
%%%%%%%%%%%%%%%%%%%%
\centerline{{\bf{Table 3}}}\\
\[\begin{array}{|lll|}\hline
J_{\pm }^1(z) e^{X^1(w)-X^2(w)} &=&  (z-w) :J_{\pm }^1(z) e^{X^1(w)-X^2(w)}
:,\\
J_{\pm }^1(z) e^{Y^1(w)}&=& {q^{\pm 1} \ov {z-wq^{\pm 1}}} :J_{\pm }^1(z)
e^{Y^1(w)}:,\\
J_{\pm }^1(z) e^{Y^2(w)}&=&:J_{\pm }^1(z) e^{Y^2(w)}:,\\
J_{\pm }^2(z) e^{X^1(w)-X^2(w)} &=& q^{k+4} (z-wq^2) (z-w) (z-wq^{-2})
:J_{\pm }^2(z) e^{X^1(w)-X^2(w)} :,\\
J_{\pm }^2(z) e^{Y^1(w)}&=&  { 1 \ov { (z-wq)(z-wq^{-1})}}
:J_{\pm }^2(z) e^{Y^1(w)}:,\\
J_{\pm }^2(z) e^{Y^2(w)}&=&  q^{k+2\pm 1} (z-wq^{\mp 1})
:J_{\pm }^2(z) e^{Y^2(w)}:.\\ \hline
\end{array}\]
%%%%%%%%%%%%%%%%%%%%%
\newpage
\centerline{{\bf{Table 4}}}\\
\[\begin{array}{|rll|}\hline
J_{\pm}^3(z) e^{X^1(w)}&=& { 1\ov {z-wq}} :J_{\pm}^3(z) e^{X^1(w)}:\\
J_{\pm}^3(z) e^{X^2(w)}&=& q^{\pm(k+3)} (z-wq^{k+2\mp(k+3)}):J_{\pm}^3(z)
e^{X^2(w)}:\\
J_{\pm}^3(z) e^{Y^1(w)}&=&(z-w):J_{\pm}^3(z) e^{Y^1(w)}:\\
J_{\pm}^3(z) e^{Y^2(w)}&=& {q^{\mp(k+2)}\ov {z-wq^{k+2\mp (k+2)}}}
:J_{\pm}^3(z) e^{Y^2(w)}:\\
%%%%%%%%%%%%%%%%%
J_{\pm}^4(z) e^{X^1(w)}&=&q^{-1-k\pm k} { (z-wq^{1+k\mp k} )\ov
(z-wq)}:J_{\pm}^4(z) e^{X^1(w)}:\\
J_{\pm}^4(z) e^{X^2(w)}&=&q^{-2}:J_{\pm}^4(z) e^{X^2(w)}:\\
J_{\pm}^4(z) e^{Y^1(w)}&=& {q^{k+1\mp (k+1)} (z-w) \ov (z-wq^{k+1\mp (k+1)})}
:J_{\pm}^4(z) e^{Y^1(w)}:\\
J_{\pm}^4(z) e^{Y^2 (w)}&=& :J_{\pm}^4(z) e^{Y^2 (w)}:\\
%%%%%%%%%%%%%%%%%
J_{\pm}^5(z) e^{X^1(w)}&=& {q^{\pm k}\ov (z-wq^{\pm k})}  :J_{\pm}^5(z)
e^{X^1(w)}:\\
J_{\pm}^5(z) e^{X^2(w)}&=&:J_{\pm}^5(z) e^{X^2(w)}:\\
J_{\pm}^5(z) e^{Y^1(w)}&=& q^{\mp (k+1)} (z-w^{\pm (k+1)}):J_{\pm}^5(z)
e^{Y^1(w)}:\\
J_{\pm}^5(z) e^{Y^2 (w)}&=& :J_{\pm}^5(z) e^{Y^2 (w)}:\\
%%%%%%%%%%%%%%%%%
J_{\pm}^6(z) e^{X^1(w)}&=& q^{1-k} (z-wq^{k-1}):J_{\pm}^6(z) e^{X^1(w)}:\\
J_{\pm}^6(z) e^{X^2(w)}&=& {q^{-2}\ov (z-wq^{k+1})}:J_{\pm}^6(z) e^{X^2(w)}:\\
J_{\pm}^6(z) e^{Y^1(w)}&=& { q^k\ov {z-wq^k}} :J_{\pm}^6(z) e^{Y^1(w)}:\\
J_{\pm}^6(z) e^{Y^2 (w)}&=&q^{1\pm 1} (z-wq^{k+1\mp 1}):J_{\pm}^6(z) e^{Y^2
(w)}:\\
%%%%%%%
J_{\pm}^7(z) e^{X^1(w)}&=& :J_{\pm}^7(z) e^{X^1(w)}:\\
J_{\pm}^7(z) e^{X^2(w)}&=& q^{\mp 1} { (z-wq^{k+3})\ov {(z-wq^{k+2\mp 1})}}
:J_{\pm}^7(z) e^{X^2(w)}:\\
J_{\pm}^7(z) e^{Y^1(w)}&=& :J_{\pm}^7(z) e^{Y^1(w)}:\\
J_{\pm}^7(z) e^{Y^2 (w)}&=& :J_{\pm}^7(z) e^{Y^2 (w)}:\\
\psi^i_\pm(zq^{\pm k/2})e^{\pm(X^1(w)-X^2(w))}&=&
:\psi^i_\pm(zq^{\pm k/2})e^{\pm(X^1(w)-X^2(w))}:,\ws i=1,2;\\
\psi^i_\pm(zq^{\pm k/2})e^{\pm Y^1(w)}&=&:\psi^i_\pm(zq^{\pm k/2})
e^{\pm Y^1(w)}:,\ws i=1,2;\\
\psi^i_\pm(zq^{\pm k/2})e^{\pm Y^2(w)}&=&:\psi^i_\pm(zq^{\pm k/2})
e^{\pm Y^2(w)}:,\ws i=1,2.
\\ \hline
\end{array}\]
%%%%%%%%%%%%%%%%%%%%%%%%%%%%%%%%%%%%%%%%%
\pagebreak

%%%%%%%%%%%%%%%%%%%%%%%%%%%%%%%%%%%%%%%%
\end{document}